\documentclass[prb,showpacs,floatfix,preprint]{revtex4}

\usepackage{epsfig} \usepackage{amsmath}

\def\cco{CoCr$_2$O$_4$}  \def\Tc{$T_c$}

\begin{document}

\title{Dielectric anomalies and spiral magnetic order in \cco} 

\author{G.\,Lawes$^1$, B.\,Melot$^2$, K.\,Page,$^2$ C.\,Ederer,$^2$ 
M.\,A.\,Hayward,$^3$ Th.\,Proffen$^4$ and  R.\,Seshadri$^2$} 

\affiliation{$^1$  Department of Physics and Astronomy, Wayne State University,
Detroit, MI 48201}  \affiliation{$^2$ Materials Department and Materials
Research Laboratory\\ University of California, Santa Barbara, CA 93106}
\affiliation{$^3$  Department of Chemistry, Inorganic Chemistry Laboratory,\\
University of Oxford, South Parks Road, Oxford, OX1 3QR UK} \affiliation{$^4$ 
Los Alamos National Laboratory, Manuel Lujan Jr. Neutron Scattering Center\\ 
LANSCE-12, MS H805, Los Alamos, NM 87545}

\date{\today}

\begin{abstract}  We have investigated the structural, magnetic, thermodynamic,
and dielectric properties of polycrystalline \cco, an insulating spinel
exhibiting both ferrimagnetic and spiral magnetic structures.  Below  \Tc =
94\,K the sample develops long-range ferrimagnetic order, and we attribute a
sharp phase transition at $T_N$ $\approx$ 25\,K with the onset of long-range
spiral  magnetic order. Neutron measurements confirm that while the structure 
remains cubic at 80\,K and at 11\,K; there is complex magnetic ordering by 
11\,K. Density functional theory supports the view of a ferrimagnetic 
semiconductor with magnetic interactions consistent with non-collinear 
ordering. Capacitance measurements on \cco\, show a sharp decrease in the
dielectric constant at $T_N$, but also an anomaly showing thermal hysteresis
falling between approximately $T$ = 50\,K and $T$ = 57\,K.   We tentatively
attribute the appearance of this higher temperature dielectric anomaly to the
development of \textit{short-range} spiral magnetic order, and discuss these
results in the context of utilizing dielectric spectroscopy to investigate
non-collinear short-range magnetic structures. \end{abstract}


\maketitle

\section{Introduction}

In recent years there has been renewed interest in investigating systems in
which charge and spin degrees of freedom are strongly coupled.  From colossal
magnetoresistive materials\cite{CMR} to diluted magnetic
semiconductors\cite{DMS}, studies on  systems exhibiting an interplay between
their magnetic and electronic properties have led to a deeper appreciation of
the role of spin-charge coupling in determining the behavior of materials.
The long-standing problem  of understanding magnetodielectric coupling in
insulating magnetic compounds has also been re-visited lately.  Recent results
on magnetodielectrics have  highlighted several important features in these
systems, including the  observation that incommensurate non-collinear magnetic
structures tend to lead to large magnetocapacitive 
couplings,\cite{GotoPRL,KimuraNature} as does
geometrical magnetic frustration. With this background, it is natural to ask how
one can use dielectric  measurements to extract information about incommensurate
non-collinear  magnetic structures in frustrated magnets.  

\begin{figure} \epsfig{file=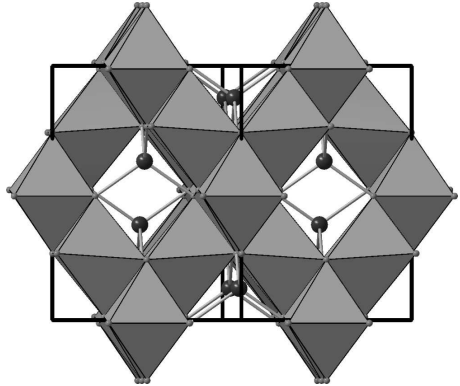, width=6cm} \caption{Structure projected
down [110] showing octahedra around Cr and  tetrahedral Co atoms.} \end{figure}

\cco\, is a spinel ferrimagnet, with Co$^{2+}$ ions on the A sites and Cr$^{3+}$
on the B sites of the spinel structure (FIG.\,1).  Both Co$^{2+}$ and Cr$^{3+}$
are magnetic; in conjunction with the fact that  coupling between the B sites is
also quite strong, a complex magnetic phase  diagram emerges. Below
approximately \Tc = 94\,K the system has long-range  ferrimagnetic order.  There
remain some discrepancies concerning the  low temperature magnetic phase
diagram.  In the original investigation on  powder \cco\, samples, Menyuk {\it
et al.} found evidence for short-range  spiral magnetic order below $T$ $\sim$
86\,K, and long-range spiral magnetic order below 
$T$ $\sim$ 31\,K.\cite{menyuk} Motivated by
suggestions that ferrimagnetic spiral long-range order in \cco\, should be
unstable, Tomiyasu {\it et al.} investigated both \cco, and MnCr$_2$O$_4$
through extensive  neutron and magnetization measurements on single crystals. 
These more recent  studies suggest that the spiral component develops
incommensurate short-range  order below approximately $T$ = 50\,K, and that this
short-range order  persists to the lowest temperatures with a correlation length
of only  3.1\,nm at $T$ = 8\,K. The authors attribute this behavior to weak
geometric  magnetic frustration.

These properties suggest that \cco\, would be a promising material for
investigations on magnetodielectric coupling in systems with non-collinear
magnetic order.  \cco\, is insulating, and has two different types of magnetic 
structures
(ferrimagnetic and spiral).  Measuring  the dielectric constant in phases with
both the ferrimagnetic and spiral components will offer insight into how
dielectric properties couple to different magnetic structures within the same
material.  

\section{Experimental and computational details}

Ceramic samples (beautiful jade-green pellets) were prepared from cobalt oxalate
[Co(C$_2$O$_4$)$_2$$\cdot$2H$_2$O] and chromium oxide (Cr$_2$O$_3$) by heating
well-ground mixtures initially at 800$^\circ$C for close to 24\,h followed by
regrinding, pelletizing and heating at 1000$^\circ$C for 24\,h. The samples were
initially characterized by lab powder x-ray diffraction  (Scintag X-II,
CuK$\alpha$ radiation) and then by time-of-flight powder neutron diffraction at
room temperatures, 80\,K and 11\,K  at the neutron powder diffractometer
(NPDF)\cite{Proffen} at the Lujan Center at Los Alamos National Laboratory. For
the measurements, the samples were contained in vanadium cans.

Calculations were performed using the projector augmented wave \cite{paw} method
implemented in the Vienna Ab-initio Simulation Package
(\textsc{vasp}).\cite{vasp} The cubic spinel structure (space group $Fd\bar 3m$)
with the lattice parameters fixed to the experimentally  obtained $a$ =
8.3351\,\AA\/ was used, with an internal structural parameter  for the oxygen of
$x$ = 0.264 determined from the diffraction data. Different collinear magnetic
structures were calculated in order to extract the Heisenberg exchange
constants. The plane wave energy cut-off was 400\,eV and we used a 8 $\times$ 8
$\times$ 8 Monkhorst-Pack mesh of $k$-points.  The LDA + $U$ method of Dudarev
\textit{et al.}\cite{dudarev} was used and $U_\text{eff}$ was varied
independently on both the Co and Cr sites.

\section{Results and discussion}

We measured the temperature and field dependence of the DC magnetization  of
\cco\, using a Quantum Design MPMS SQUID magnetometer, operating between  $T$ =
5\,K and $T$ = 350\,K, and measured the AC magnetization of the sample on a
Quantum Design Physical Property Measurement System PPMS at  frequencies between
$\omega/2\pi$ = 100\,Hz  and $\omega/2\pi$ = 10\,kHz. We measured the specific
heat of \cco\, as a function of temperature and magnetic field with a
quasi-adiabatic method on the PPMS using a 20\,mg pressed powder pellet.  We
found that the internal time constant of the sample was significantly smaller
than the other time constants in the fit, suggesting excellent thermal coupling
between the sample and calorimeter.  In order to measure the dielectric
constant, we pressed approximately 75 mg of \cco\, into a  5\,mm $\times$ 5\,mm
$\times$ 1.5\,mm pellet, and fashioned electrodes on  opposite faces of the
pellet using conducting silver epoxy. We measured the  dielectric constant
between $\omega/2\pi$ = 1\,kHz and $\omega/2\pi$ = 1\,MHz  with an Agilent 4284A
LCR meter, using the PPMS for temperature and  magnetic field control. The data
at lower frequencies was qualitatively similar to the higher frequency
measurements, although they were somewhat noisier and  displayed larger losses,
possibly due to conduction at the grain boundaries.

\begin{figure} \epsfig{file=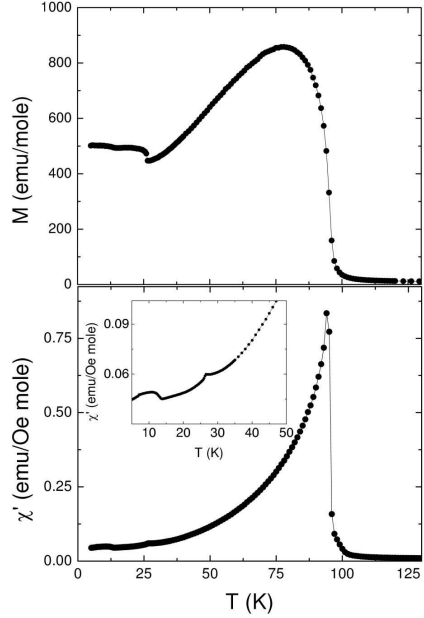, width=6cm} \caption{Upper panel: 
Field-cooled DC magnetization of \cco\, measured in an applied field of H=100
Oe. Lower Panel:  Real part of the AC susceptibility of \cco\, measured at H=0
and $\omega/2\pi$=10 kHz.  Inset: Lower temperature magnetic anomalies in \cco.}
\end{figure}

Figure\,2 shows the magnetization as a function of temperature measured on
warming after cooling in a magnetic field (upper panel).  These data are
qualitatively similar to the FC data obtained by  Tomiyasu {\it et
al.}\cite{tomiyasu}   The magnetization shows a large increase below the onset
of ferrimagnetic  long range order at \Tc = 95\,K, and there is an anomaly at
$T_S$ = 27\,K associated with spiral magnetic order in the system.  It should be
noted that the anomaly at $T_S$ = 27\,K is sharper than that presented in the Tomiyasu
{\it et al.} data.\cite{tomiyasu}  

The lower panel of Fig.\,2 plots the AC magnetization measurements at 10 kHz. 
The real part of the susceptibility exhibit a very large peak at \Tc. The inset
to the lower panel of Fig\,2 shows the low temperature AC susceptibility in more
detail.  There is a very clear peak in $\chi'$ at $T_S$ = 27\,K, consistent with
the sharp anomaly in the DC magnetization measurements. Furthermore, there is
another clear magnetic feature at approximately $T$ = 13\,K, which is  rather
more difficult to observe in the DC measurements. Tomiyasu {\it et al} 
associate this feature with a saturation of the correlation  length for the
spiral component on cooling.\cite{tomiyasu} These bulk  magnetization
measurements conclusively show the existence of magnetic  features at \Tc =
95\,K and T$_S$ = 27\,K, associated with the ferrimagnetic  and spiral magnetic
structures respectively, and a third magnetic feature at  $T$ = 13\,K.

\begin{figure} \epsfig{file=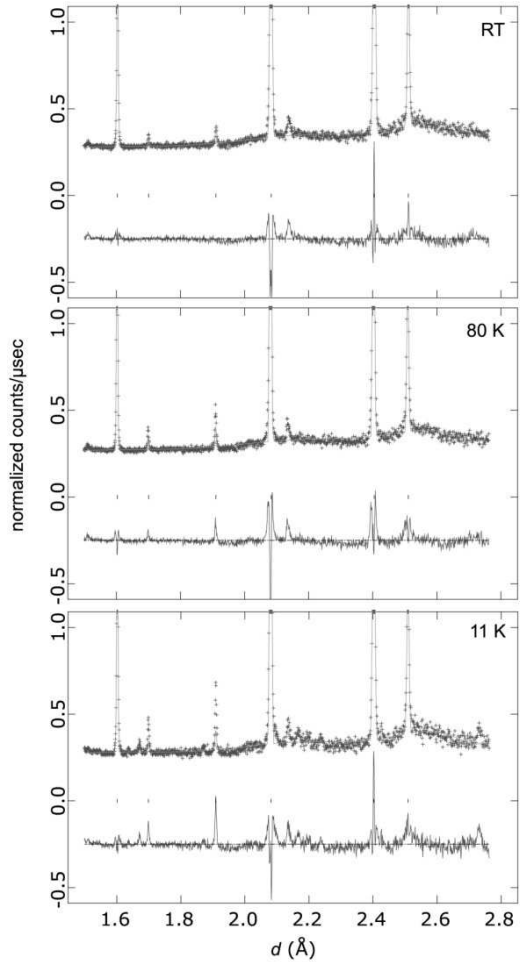, width=7cm} \caption{Time-of-flight neutron
powder diffraction at room temperature,  80\,K and 11\,K. The symbols in each
panel are data, the lines are fits  using the Rietveld method to the nuclear
structure, and the traces at the  bottom of each panel are the difference
profiles. The peaks in the difference profiles that emerge as the temperature is
lowered arise from long-range magnetic ordering.} \end{figure}

In order to get more information about the magnetic transitions in \cco, we also
conducted neutron measurements on these samples.  Figure\,3 displays the results
of fitting the low-angle time-of-flight  neutron diffraction data from \cco\/
powders to the \textit{nuclear\/} structure. At room temperature, the data is
almost completely fit except for a small reflection near $d$ = 2.15\,\AA\/ which
corresponds to the sample container. At 80\,K, below the first collinear
magnetic ordering temperature, we find that some of the peaks whose nuclear
contribution is fit, have a  residual intensity seen clearly in the difference
spectra, particularly  near $d$ = 1.7\,\AA\/ and 1.9\,\AA, due to magnetic
ordering. At 11\,K, the data reveal a number of peaks whose intensities are not
correctly fit, as well as new reflections which have come about because of the
long-range non-collinear magnetic order, again evident in the difference
spectrum. In particular, the low-angle peak at 2.7\,\AA\/ is evident only in the
11\,K data as are peaks near $d$ = 1.7\,\AA\/ and $d$ = 2.2\,\AA. While we have
not at this stage been able to completely establish the magnetic structure, we 
present these data to strengthen our interpretation of the magnetic and 
transport measurements.

The electronic structure of \cco\/ was calculated using the LSDA + $U$
methodology. FIG.\,4 displays the total density of states (DOS) and the partial
densities of Co $d$ and Cr $d$ states obtained using $U_\text{eff}^\text{Co}$ =
$U_\text{eff}^\text{Cr}$ = 0\,eV (LSDA, top panel), as well as
$U_\text{eff}^{Co}$ = 4\,eV and $U_\text{eff}^\text{Cr}$ = 2\,eV (lower panel)
and a collinear N{\'e}el-type magnetic structure (Co spins antiparallel to Cr
spins). The system is an insulating fully spin-polarized ferrimagnetic
semiconductor already in LSDA, albeit with a very small gap, less than 0.1\,eV. 
Application of $U$ on either of the two magnetic sites enlarges the gap but only
application of $U$ on both sites leads to a significant gap of $>$ 1\,eV. The
calculations show that this system can be expected to be a robust insulator and
therefore promising for magnetodielectric measurements.

\begin{figure} \epsfig{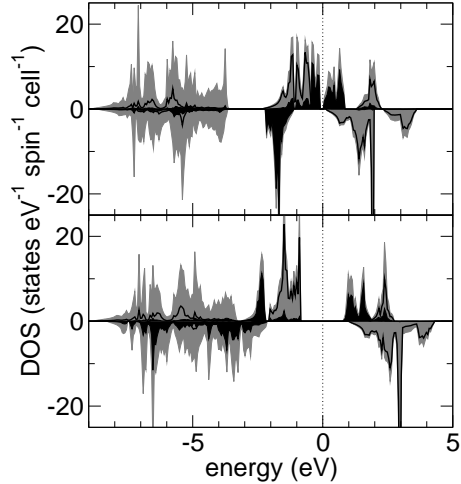} \caption{Total densities of
state (gray shaded) and partial densities of Co $d$ (black shaded) and Cr $d$
(black line) states obtained for $U_\text{eff}^\text{Co}$ =
$U_\text{eff}^\text{Cr}$ = 0\,eV (LSDA, top panel), as well as
$U_\text{eff}^{Co}$ = 4\,eV and $U_\text{eff}^\text{Cr}$ = 2\,eV (lower panel)
and a collinear N{\'e}el-type ordering.} \end{figure}

We also extracted the Heisenberg exchange coupling constants by calculating the
total energy differences for different collinear magnetic configurations. It has
been shown by Kaplan \cite{Kaplan} that using a simple Heisenberg model for the
cubic spinel structure with only nearest neighbor couplings $J_{AB}$ and
$J_{BB}$, the magnetic ground state structure is determined by the parameter $u
= (4 J_{BB} S_B)/(3 J_{AB} S_A)$, where $S_A$ and $S_B$ denote the spins of the
$A$ and $B$-site cations respectively. The collinear ferrimagnetic N{\'e}el
state is the stable ground state for $u < u_0 = 8/9$, whereas for larger values
of $u$ non-collinear spiral ordering is expected. The Heisenberg exchange
constants extracted from our calculations, using the reasonable values
$U_\text{eff}^{Co}$ = 4\,eV and $U_\text{eff}^\text{Cr}$ = 2\,eV, are: $S_A
J_{AB} S_B$ = 5.0\,meV, $S_B J_{BB} S_B$ = 2.4\,meV, and $u$ = 0.65. Increasing
$U_\text{eff}^\text{Co}$ and decreasing $U_\text{eff}^\text{Cr}$ leads to a
larger value of $u$, eventually exceeding the critical value $u_0 = 8/9$. In
addition, we also extracted the coupling constant $J_{AA}$ which is neglected in
the treatment of Kaplan. For the same values of $U$ as above
($U_\text{eff}^{Co}$ = 4\,eV and $U_\text{eff}^\text{Cr}$ = 2\,eV) we obtain
$S_A J_{AA} S_A$ = 0.7\,meV. This value is not necessarily negligible and
indicates that the coupling between the A-site cations could play an important
role in the magnetic properties of cubic spinels and in fact lead to
non-collinear magnetic order even for values of $u$ smaller than $u_0=8/9$. The
possible importance of A-site coupling has also been pointed out in a recent
experimental study of the spinel systems $M$Al$_2$O$_4$ ($M$=Co, Fe, Mn)
\cite{Loidl}. A detailed analysis of the $U$ dependence of the electronic
structure and magnetic couplings in CoCr$_2$O$_4$ is in progress. From the
preliminary results presented in this work it can be concluded that the
calculated values of the exchange coupling constants are consistent with the
appearance of non-collinear magnetic order and that $A$ site coupling might be
important to understand the magnetic properties of this system.

In an attempt to establish whether the spiral magnetic transition at 
$T_S$ = 27\,K  is long range\cite{menyuk} or short range\cite{tomiyasu}, we measured
the  specific heat of our polycrystalline \cco\, sample.  The thermodynamic
data  presented in FIG.\,5, including both the lattice and magnetic
contributions  to heat capacity, strongly suggests that there are two
transitions to magnetic states with long range order, one occurring with \Tc =
94\,K and the second at  $T_S$ = 27\,K.  Specifically, the sharp peak in $C/T$ at
$T_S$ = 27\,K is  indicative of long range collective ordering.  We can calculate
the entropy  under the peaks in the upper panel of FIG.\,5 to estimate the
fraction of  spins participating in the long range order at each transition. 
The entropy lost by \cco\, at \Tc\, is approximately 0.1\,J/K per mole, 
 while the entropy change at $T_S$ = 27\,K is approximately  0.15\,J/K
per mole, both at $H$ = 0.  These changes in entropy are  small
compared to the expected change in entropy for fully spin  ordered \cco\, of
34.6 J/K per mole (using the spin-only values for Co$^{2+}$ and
Cr$^{3+}$).  This is consistent with the relatively large specific heat
exhibited by \cco\, over the entire temperature range shown, which suggests that
significant short range magnetic order is likely developing over a wide range of
temperatures.  We do not see any features in specific heat which could be
associated with a thermodynamic transition occurring at $T$ = 13\,K; this low
temperature anomaly in the magnetization does not involve any measurable change
in entropy of the system.

\begin{figure} \epsfig{file=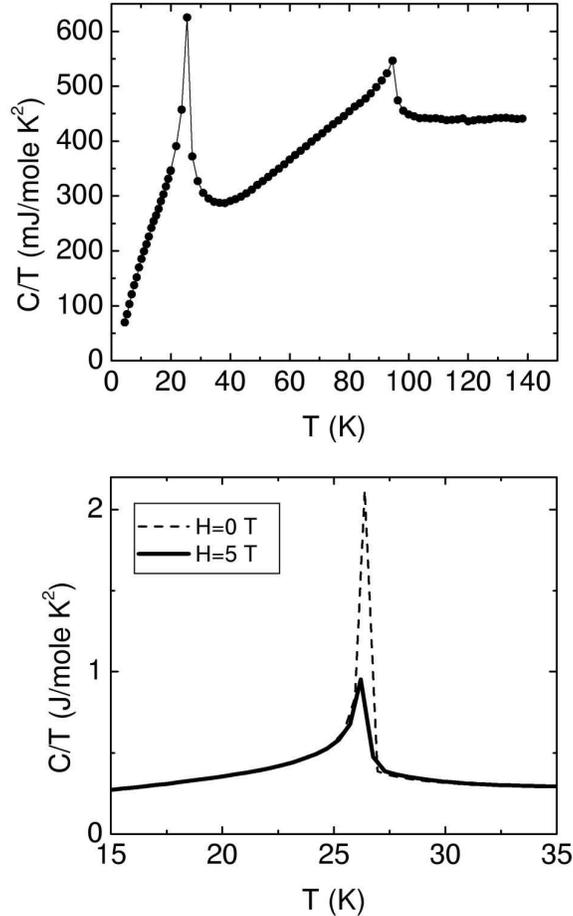, width=8cm} \caption{Upper Panel:  Specific
heat measurements on \cco\, plotted as $C/T$ vs $T$.  Lower Panel:  Specific
heat of \cco\, at the  $T_S$ = 27\,K transition, measured at $H$ = 0\,T and $H$ =
5\,T.} \end{figure}

The lower panel of FIG.\,5 plots the dependence of the $T_S$ = 27\,K transition at
two different magnetic fields. The $H$ = 0 peak appears to be  somewhat larger
than that plotted in the upper panel. This is because the  higher density of
data points more accurately captures the proper shape of the  peak.  The amount
of entropy under the peak is practically unchanged. The  transition temperature
is almost completely independent of applied magnetic  field, but the magnitude
of the peak is suppressed by over a factor of two at  $H$ = 5\,T as compared to
the zero field data.  This suggests that fewer  spins undergo long-range spiral
order at $T_S$ = 27\,K at larger magnetic  fields, but it is impossible to tell
from the data if the extra entropy is  removed at higher or lower temperatures. 

One of the primary motivations for  these experiments was to investigate what
effect the ferrimagnetic and spiral  magnetic transitions would have on the
dielectric constant of \cco.  These data  are plotted in the upper and middle
panels FIG.\,6, which shows the  dependence of the dielectric constant on
temperature for two different  polycrystalline \cco\, samples measured on
warming.  In both these samples,  the dielectric constant depends only weakly on
temperature, with sample B  showing a variation of approximately 1\% over the
entire temperature range,  and sample A showing an even smaller change.
Differences in the magnitude of  the dielectric constant can be attributed to
errors in determining the  geometrical capacitance factor. Differences in the
temperature dependence  may possibly arise from small differences in
conductivity between the two  samples.  Preliminary measurements suggest that
sample B is slightly more  conducting than sample A, although this needs to be
investigated in more  detail. Both samples were structurally identical (verified
by powder x-ray  diffraction) but sample B was better sintered. 

\begin{figure} \epsfig{file=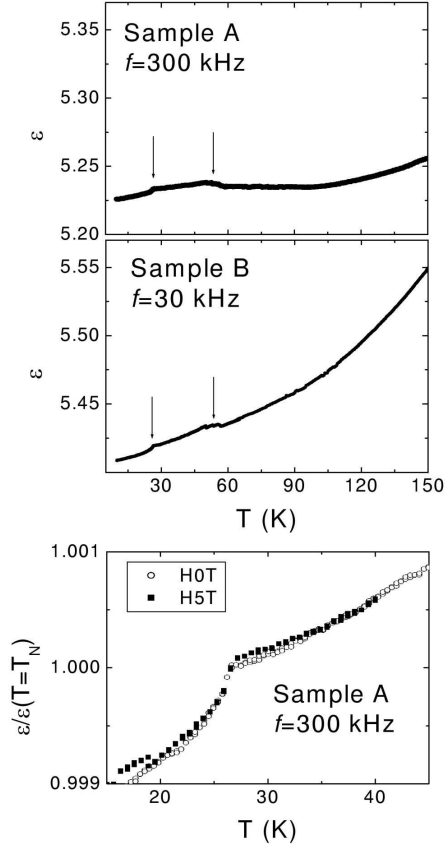, width=6cm} \caption{Top Panel:  Dielectric
constant of \cco\, sample A measured at H=0 as a function of temperature
measured at $\omega/2\pi$=300 kHz.  Middle Panel:  Dielectric constant of \cco\,
sample B measured at H=0 as a function of temperature measured at
$\omega/2\pi$=30 kHz.  The arrows indicate the location of the observed
dielectric anomalies.  Bottom Panel:  Relative dielectric constant of \cco\,
sample A at the spiral ordering transition at $H$=0 and at $H$=5 T.}
\end{figure}

The general features in the dielectric constant are very similar for both these
samples.  Most notably, there is a well-defined drop in dielectric constant
coincident with the onset of long range spiral magnetic order at $T_S$ = 27\,K
(indicated with an arrow).  This sharp decrease is a signature of the coupling
between the dielectric properties and magnetic structure.  The change in
dielectric constant is fairly small, only 0.2\% between $T_S$ = 27\,K and the 
base temperature. This should be compared with EuTiO$_3$\cite{katsufuji} and
SeCuO$_3$\cite{lawes} which show larger changes on ordering (3.5\% and 2\%
respectively).  The small magnitude of the shift in the dielectric constant
\cco\, at the spiral ordering transition may result from the the fact that only
a small fraction of spins order at this temperature, as evidenced by the
thermodynamic data discussed previously.   This dielectric anomaly is almost
completely independent of magnetic field as well (lower panel of FIG.\,6).  An
applied magnetic field of $H$ = 5\,T introduces almost no shift in either  the
temperature or magnitude of the anomaly. A magnetic field does however 
introduce a small shift in the overall dielectric constant, on the order of 
0.01\% in a field of $H$ = 1\,T (not shown). 

There is no clear dielectric feature at the ferrimagnetic ordering temperature
\Tc. There may be a subtle change of slope observable in Sample A, but this is
not seen in Sample B, which shows much larger intrinsic dependence of dielectric
constant on temperature.  The lack of a sharp anomaly at \Tc\, suggests that
there is only weak coupling between the dielectric constant of \cco\, and the
ferrimagnetic component of the magnetic structure.  Furthermore, there is no
dielectric feature associated with the $T$ = 13\,K transition. Since this 
magnetic feature is believed to be related to the spiral  magnetization
component\cite{tomiyasu}, one  would expect to see a clear anomaly at 13\,K. 
However, specific heat measurements show that only a small number of spins are
likely to be involved in this transition, which perhaps explains the absence of
any dielectric feature.

The dielectric constant of \cco\, couples only to the spiral spin structure, and
not to the ferrimagnetic structure.  The shift in dielectric constant observed
in many magnetodielectric materials is believed to arise from strong spin-phonon
coupling\cite{katsufuji,lawes}.  Within this framework, the phonon mode giving
rise to this magnetodielectric behavior must not couple to magnetic order along
the (001) direction, but should be sensitive to spiral spin-ordering with a
propagation vector along the $Q$=($\delta$, $\delta$,0) direction in the (001)
plane\cite{tomiyasu}.  This observation provides an important symmetry
restriction on allowed forms for the spin-lattice coupling in \cco.  Further
investigations on the microscopic mechanisms for spin-phonon coupling in \cco\,
will rely on a detailed understanding of the phonon modes in this system.  

We find evidence for one further temperature dependent dielectric anomaly in
\cco. The dielectric constant shows a distinct peak at approximately  $T$ =
50\,K for the two samples (indicated by arrows in FIG. \,6). More precisely, 
measurements on Sample B seem to indicate the presence of a series of peaks, 
while Sample A rather exhibits one broader peak. The magnitude of the increase 
in dielectric constant at this temperature is comparable to the decrease in 
dielectric constant below $T_S$ = 27\,K.  However, the origins of this feature 
are less clear.  Neither the magnetization data in FIG.\,2 nor the specific 
heat data in FIG.\,6 show any evidence for a phase transition at this 
temperature.  Additionally, this anomaly has a rather broad onset, which  varies
somewhat with sample history.  Since \cco\, does not have any  structural phase
transitions at low temperatures, this anomaly must have a  magnetic origin. 
Specifically, we believe that this dielectric anomaly arises  from the onset of
short range spiral magnetic order, which has also been  shown to occur at
approximately $T$ = 50\,K\cite{tomiyasu}.

As a further probe of this anomaly, we measured the dielectric constant on
warming and cooling at different rates.  FIG.\,7 plots the dielectric constant
of \cco\, measuring on warming and cooling at 1\,K/min. (upper panel) and 
5\,K/min. (lower panel). In both cases, there is {\it no} anomaly when the 
dielectric constant is measured on cooling, but there is a clear feature in  the
dielectric constant when the sample is warmed. Furthermore, the peak 
temperature for this anomaly occurs at a higher temperature ($T$ = 50\,K)  when
the sample is warmed at 5\,K/min. as compared to when it is warmed at  1\,K/min.
($T$ = 48\,K).  This thermal hysteresis may arise from a  non-equilibrium
distribution of regions of short range spiral magnetic order.  Below $T_S$ =
27\,K, the thermodynamic and dielectric data suggest that \cco\,  develops long
range spiral order.  On warming, the disappearance of a  non-equilibrium
distribution of short range spiral magnetic clusters may be  responsible for the
dielectric signal at $T$ = 50\,K.  
The shift of this anomaly to lower  temperatures as the
sample is warmed more slowly supports this idea.  Conversely, when the sample is
cooled from higher temperatures, there would be no spiral magnetic clusters
present, and therefore no shift in the dielectric constant. 

In summary, we have presented extensive magnetic, thermodynamic, neutron, and
dielectric data on \cco.  These results are consistent with a phase transition
to a ferrimagnetically ordered state at $T$=94\,K and we have conclusive
evidence for a second transition to a phase with long-range spiral magnetic
order below $T_S$ = 27\,K in our polycrystalline sample.  Magnetocapacitive
measurements show that the dielectric constant of \cco\, couples strongly to the
spiral magnetic order parameter, but is insensitive to the ferrimagnetic spin
component. This limits the allowed spin-phonon coupling symmetry in this system,
and places restrictions on the possible microscopic mechanism for the observed
magnetodielectric shifts. The dielectric constant of \cco\, is also found to be
affected by the development of {\emph short-range} spiral magnetic order.  This
suggests that in materials with strong spin-phonon coupling, it may be possible
to use capacitive measurements to probe short-range magnetic correlations.

\begin{figure} \epsfig{file=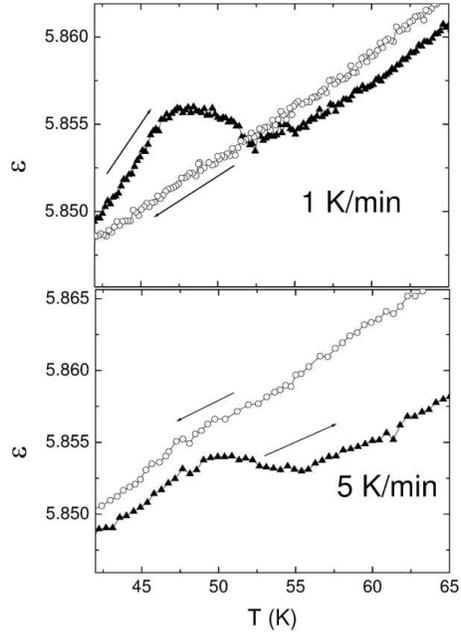, width=6cm} \caption{Upper Panel: 
Dielectric constant of \cco\, near $T$=50 K measured on cooling and warming at 1
K/min.  Lower Panel:  Dielectric constant of \cco\, near $T$=50 K measured on
cooling and warming at 5 K/min.  In both plots, the arrows indicate the
temperature change.} \end{figure}

\section*{Acknowledgements}
This work in UCSB was supported by National Science Foundation through the MRL 
program (DMR00-80034), through a Chemical Bonding Center (CHE04-34567).
Measurements at the Lujan Center at Los Alamos Neutron Science Center were
supported by the Department of Energy Office of Basic Energy Sciences and
Los Alamos National Laboratory funded by Department of Energy under contract
W-7405-ENG-36. MAH thanks the Royal Society for funding.

\clearpage

\end{document}